\documentclass[12pt]{article}

\begin{document}
\begin{center}
{\bf Vacuum birefringence caused by arbitrary spin particles}\\
\vspace{5mm}
 S. I. Kruglov  \\
\vspace{5mm}
\textit{University of Toronto at Scarborough,\\ Physical and Environmental Sciences Department, \\
1265 Military Trail, Toronto, Ontario, Canada M1C 1A4}
\end{center}

\begin{abstract}
We study the propagation of a linearly polarized laser beam in the
external transverse magnetic field taking into consideration the
vacuum polarization by arbitrary spin particles. Induced ellipticity
of the beam are evaluated using the effective Lagrangian. With the
help of the PVLAS experimental data, we obtain bounds on masses of
charged higher spin particles contributed to ellipticity.
\end{abstract}

The new PVLAS results \cite{Zavattini} indicate that there is no the
previously reported rotation of the polarization axis of the beam.
The PVLAS collaboration found no rotation and ellipticity with $2.3$
T external transverse magnetic field giving corresponding bounds on
them. It was concluded also that the ellipticity peak at $5$ T
magnetic field might be of instrumental origin.

Within particle physics, the effect of the rotation and ellipticity
can be explained by the existence of a new axion-like (spin-0)
particle (ALP) and/or minicharged particles (MCPs) \cite{Sikivie},
\cite{Sikivie1}, \cite{Maiani}, \cite{Raffelt}, \cite{Gies}, \cite{Ahlers},
\cite{Masso}, \cite{Brax}. Vacuum birefringence appears also in
Lorentz violating electrodynamics \cite{Kostelecky} and generalized
non-commutative electrodynamics \cite{Kr}. It is reasonable also to
look for alternative scenarios.

In this letter, we study vacuum birefringence caused by higher spin
particles. In \cite{Kruglov1}, \cite{Kruglov2} the non-linear
corrections to the Maxwell Lagrangian were calculated, taking into
consideration the vacuum polarization of arbitrary spin particles
possessing the electric dipole moment (EDM) and anomalous magnetic
moment (AMM). Due to the small contribution of EDM of particles
(violating the CP invariance) to physical quantities, we omit EDMs
(putting $\sigma =0$). The effective Lagrangian is given by
\footnote{We use here the standard notation for the magnetic
induction field $B_i=(1/2)\epsilon_{ijk}F_{jk}$ ($F_{\mu\nu}$ is the
electromagnetic field strength tensor) instead of $H_i$ used in
\cite{Kruglov1}.}
\begin{equation}
{\cal L}_{eff}=\frac{1}{2}\left(\textbf{E}^2-\textbf{B}^2 \right) +
a \left(\textbf{E}^2-\textbf{B}^2 \right)^2 +
b\left(\textbf{E}\textbf{B}\right)^2 , \label{1}
\end{equation}
\[
a=\frac{\epsilon \alpha ^2(2s+1)}{360m^4}\left[ s(s+1)(3s^2+3s-1)
g^4 -10s(s+1)g^2+7\right] ,
\]
\vspace{-7mm}
\begin{equation}  \label{2}
\end{equation}
\vspace{-7mm}
\[
b=\frac{\epsilon \alpha ^2(2s+1)}{90m^4}\left[1 - s(s+1)(3s^2+3s-1)
g^4 \right] ,
\]
where $\alpha =e^2/(4\pi )\simeq1/137$, $\epsilon =1$ for bosons and
$\epsilon =-1 $ for fermions, $g$ is the gyromagnetic ratio for
particles possessing the spin $s$ and the mass $m$. The rationalized
(Heaviside-Lorentz) units and $\hbar =c=1$ are used here. The
effective Lagrangian (1) generalizes the Heisenberg-Euler Lagrangian
\cite{Heisenberg} on the case of arbitrary spin particles in the
approach of \cite{Hurley}, \cite{Hurley1}, \cite{Hurley2}. At $s=1/2$ 
and $g=2$, Eq.(1) is converted into the Heisenberg-Euler Lagrangian 
corresponding to QED.

For the case of the plane electromagnetic wave
($\textbf{e},\textbf{b}$) traveling in $z$-direction, the components
of the polarization vector are given by
\begin{equation}
e_\bot=E_0\sin\theta\exp i\left(k_\bot z-\omega
t\right),~~~~e_\|=E_0\cos\theta\exp i\left(k_\| z-\omega t\right),
\label{3}
\end{equation}
where $k_\bot=n_\bot \omega$, $k_\|=n_\| \omega$; $n_\bot$, $n_\|$
are indexes of refraction corresponding to the electric field of the
plane wave $\textbf{e}$ perpendicular and parallel to the background
magnetic induction field $\overline{\textbf{B}}=(\overline{B},0,0)$.
The $\theta$ is the angle between the polarization vector
$\textbf{e}$ and the external magnetic induction field
$\overline{\textbf{B}}$.

In \cite{Kr1}, we have obtained parameters of elliptically polarized
wave, using the notations of \cite{Born}, for the electromagnetic
fields described by Lagrangian (1) for arbitrary $a$ and $b$:
\begin{equation}
\alpha=\theta,
~~\delta=\left(k_\bot-k_\|\right)z=(4a-b)\omega\overline{B}^2z,~~\sin2\chi=
\left(\sin2\alpha \right)\sin\delta . \label{4}
\end{equation}
The shape and orientation of the vibrational ellipse (ellipticity)
is given by
\begin{equation}
\Psi\equiv\tan\chi\simeq\chi\simeq
\frac{1}{2}\delta\sin2\theta=\frac{1}{2}(4a-b)\omega\overline{B}^2z\sin2\theta
, \label{5}
\end{equation}
where $\omega=2\pi/\lambda$, $\lambda$ is a wave length. When the
electromagnetic wave propagates the distance $L$, Eq.(5), with the
help of Eq.(2) becomes
\begin{equation}
\Psi=\frac{\epsilon \alpha ^2\overline{B}^2\pi L(2s+1)}{45\lambda
m^4}\left[ s(s+1)(3s^2+3s-1) g^4 -5s(s+1)g^2+3\right]\sin2\theta ,
\label{6}
\end{equation}
After traveling the distance $L$, initially linearly polarized wave
becomes elliptically polarized wave and there is no rotation of the
polarization axis of the beam. It follows from Eq.(6) that
ellipticity increases with the spin of particles.

Eq.(6) allows us to consider differing scenarios explaining
observations in the PVLAS experiment. With the help of the data
\cite{Zavattini}
\[
\Psi\leq 3.1\times10^{-13}\frac{1}{{\mbox p }{\mbox a}{\mbox
s}{\mbox s}},~~L=1~{\mbox m},
\]
\vspace{-7mm}
\begin{equation}  \label{7}
\end{equation}
\vspace{-7mm}
\[
\lambda=1064~ {\mbox n}{\mbox
m},~~\theta=\frac{\pi}{4},~~\overline{B}=2.3 ~{\mbox T} ,
\]
and using the value $e\overline{B}=1.36\times10^{-10}$ MeV$^2$ (for
$\overline{B}=2.3$ T), $g=2$, we find from Eq.(6) lower bounds on
the possible masses of particles with higher spins contributed to
ellipticity
\[
\begin{array}{ccccc}
 s & 1 & 3/2 & 2 & 5/2 \\
 m({\mbox MeV}) & 0.17 &0.26  & 0.36 &0.46
\end{array}
\]

The mass of a particle as a function of spin-$s$ presented here is a
lower bound satisfying the data (7). As the data (7) are
preliminary, the possible masses of particles contributed to
ellipticity are also preliminary. Eq. (6), may be used to fix
particle masses from experiments. Here we imply that only one kind
of particles results in the vacuum polarization. The contribution of
electrons (s=1/2) to vacuum polarization (and ellipticity) is very
small and can be neglected. It should be noted that ellipticity for
bosons and fermions have opposite signs because of the factor
$\epsilon$ in Eq.(6). For fermions, ellipticity is negative
corresponding to the left-handed polarization, and for bosons,
ellipticity is (positive) the right-handed. One may consider also
the possibility of the existence of particles with higher spins
($s>5/2$) contributed to the effective Lagrangian (and ellipticity).
However, masses evaluated are too low. Indeed, if such particles
exist, they can be produced in electron-positron collisions. It
means that the explanation of PVLAS experiment on the base of vacuum
polarization by higher spin particles require lower ellipticity.  An
improvement in the PVLAS experiment would correct masses evaluated
to the higher values.

So, based on the effective Lagrangian taking into consideration the
vacuum polarization of arbitrary spin particles, we find, from
experimental data, lower bounds on masses of higher spin particles
contributed to ellipticity.

\end{document}